\input epsf
\documentstyle[preprint,aps,pre]{revtex}     
\tighten
\begin{document}
\draft

\title{Magnetization Profile in the
$\bbox{d=2}$ Semi-Infinite Ising Model\\[4mm]
and Crossover between Ordinary and
Normal Transition\\[1cm]}
\author{Peter Czerner\footnote{e-mail: peterc@theo-phys.uni-essen.de}
 and Uwe Ritschel\footnote{e-mail: uwe@theo-phys.uni-essen.de}\vspace*{1cm}}
\address{Fachbereich Physik, Universit\"at GH Essen, 45117 Essen
(F\ R\ Germany)\\[2cm]}
\maketitle
\narrowtext
\begin{abstract}
We theoretically investigate
the spatial dependence of the order parameter of the
two-dimensional semi-infinite Ising model with a free surface
at or above the bulk critical temperature.
Special attention is paid to the influence of
a surface magnetic field $h_1$ and the crossover between
the fixed points at $h_1=0$ and $h_1=\infty$.
The sharp {\it increase}
of the magnetization $m(z)$ close to the boundary generated
by a {\it small} $h_1$, which was found by the present authors
in the three-dimensional model, is also seen
in two dimensions. There, however, the universal short-distance power law
is modified by a logarithm, $m(z)
\sim z^{\kappa}\,\mbox{ln}\,z$, where
$\kappa$, the
difference between the scaling dimensions of $h_1$ and
the bulk magnetization, has the exact value $3/8$.
By means of a phenomenological scaling analysis, the short-distance
behavior can be related to
the logarithmic dependence of the surface magnetization $m_1$
on $h_1$. Our results, which are corroborated by Monte
Carlo simulations, provide a deeper understanding
of the existing exact results
concerning the local magnetization and relate the
short-distance phenomena in two dimensions to those in higher
dimensionality.
\end{abstract}
\pacs{PACS: 75.40.-s,75.40.Mg,68.35.Rh,05.50.+q}
\section{Introduction}\label{intro}
In systems at or close to a critical point the asymptotic power laws
governing the behavior of thermodynamic
quantities are modified in the vicinity
of surfaces or
other imhomogenities\cite{binder,diehl}.
The characteristic distance within which changes
occur is given by the bulk correlation length $\xi$.
In general each bulk
universality class splits up in several {\it surface
universality classes}, depending upon whether the tendency to order
in the surface is smaller or larger than or the same as in the bulk.
In the case of the
two-dimensional (2-$d$) Ising model in a semi-infinite geometry there exist
two surface universality classes. Since the boundary is one-dimensional
and, thus cannot become critical itself, the surface generally reduces the
tendency to order. More precisely, for any positive starting value
the exchange coupling between surface spins $J_1$ is driven to zero by
successive renormalization-group transformations.
The relevant field pertaining to
the surface is the magnetic field $h_1$,
which acts on surface spins only and which
for instance may take into account the influence of an
adjacent noncritical medium. The two universality classes then are
labelled by $h_1=0$, the ``ordinary'' transition, and $h_1=\infty$, the
``normal'' transition, where the former is a unstable fixed point and
the latter is a stable fixed point of the
renormalization-group flow\cite{binder,diehl}.

In this work
we are mainly concerned with order-parameter profiles
for {\it finite} $h_1$ in the crossover region between the fixed points.
Nonetheless, let us first recapitulate the situation {\it at} the fixed points.
For the sake of simplicity, in the Introduction our
considerations remain restricted to bulk criticality $T=T_c$, but
the extension to the critical region is straightforward and
will be done below.
At the critical point and $h_1=0$, the order-parameter
(or magnetization) profile
$m(z)$ is zero
for any distance $z\ge 0$ from the surface. In the other extreme,
$h_1=\infty$,
it is well known that at macroscopic distances from the surface the
magnetization decays as $\sim z^{-x_{\phi}}$,
where $x_{\phi}=\beta/\nu$ is the scaling dimension of the bulk order-parameter
field, with the exact value $1/8$ in
the 2-$d$ Ising model.

What do we expect, when $h_1$ takes some intermediate
value, i.e., in the crossover region between the fixed points?
Now, $h_1$ will certainly generate a surface magnetization
$m_1$. As far as the profile $m(z)$ is concerned,
a first guess would perhaps be that the magnetization should
decay from that value
as $z$ increases away from the surface. This guess is supported
by mean-field theory, where one can calculate $m(z)$
and indeed finds a monotonously decreasing function of $z$
\cite{binder,lubrub,bray}.

In Ref.\,\cite{czeri} the present authors have shown that,
contrary to the naive (mean-field)
expectation, fluctuations may cause the order parameter to steeply
{\it increase} to values $m(z)\gg m_1$
in a surface-near regime. The range
within which this growth occurs (at bulk criticality) is determined
by $h_1$, the characteristic
length scale being $l_{1}\sim h_1^{-1/y_1}$, where
$y_1=\Delta_1/\nu$ is the scaling dimension of $h_1$ \cite{diehl}.
As further demonstrated by the present authors in Ref.\,\cite{czeri},
the growth of order in the near-surface regime $z\ll l_1$
is described by a {\it universal} power law
\begin{equation}\label{power}
m \sim z^{\kappa}\quad \mbox{with} \quad \kappa=y_1-x_{\phi}\>,
\end{equation}
i.e., the growth exponent $\kappa$ is governed by the difference
between the scaling dimensions $y_1$ and $x_{\phi}$.
For $z\simeq l_1$ the profile has a maximum and
farther away from the surface, at distances much
larger than $l_{1}$, the magnetization decays as $z^{-x_{\phi}}$.

Largely analogous results---monotonous behavior at the fixed points
and profiles with one extremum in the crossover regime---where
previously found
by Mikheev and Fisher \cite{mifi} for the {\it energy density}
of the 2-$d$ Ising model. The authors also suggested to calculate
the order parameter in the crossover region.
Below we focus our attention exactly on this problem.
The questions posed in this work are the following:
\begin{itemize}
\item Does the short-distance
growth of $m(z)$, found for instance in the three-dimensional Ising
model, also occur in two dimensions?
\item Does the simple power
law (\ref{power}) quantitatively describe the magnetization profile
in $d=2$, or are modifications to be expected?
\item Is the scenario for the crossover between ``ordinary'' and
``normal'' transition, developed for the three-dimensional
Ising model, also valid in $d=2$?
\end{itemize}
As we will demonstrate below, the answer to the first
and third question is ``yes",
but the simple power law (\ref{power}) is modified by a {\it logarithm}.

The rest of this paper is organized as follows: In Sec.\,\ref{two}
the theoretical framework is expounded.
We first summarize the results of Ref.\,\cite{czeri} and
then generalize the scaling analysis
by taking into account the available
exact results on the dependence of $m_1$ on $h_1$ and
on the magnetization profile.
In Sec.\,\ref{three}, in order to corroborate our analytical findings,
we present Monte Carlo (MC) data for $m(z)$ and
compare with exact results for the order-parameter profile.
In two Appendices exact literature results on the dependence
of the surface magnetization on $h_1$ and on the order-parameter
profile are briefly reviewed.

\section{Theory}\label{two}
We consider the semi-infinite Ising system with a free boundary
on a plane square lattice.
The exchange coupling between
neighboring spins is $J$.
A surface magnetic field $H_1$ is
imposed on the boundary spins and bulk magnetic fields are set to
zero such that the Hamiltonian of the model reads
\begin{equation}\label{ising}
{\cal H}= -J\!\sum_{<ij>\in V}\,s_is_j-H_1\sum_{i\in\partial V}\,s_i\>,
\end{equation}
where
$\partial V$ and $V$ stand for the boundary and the
whole system (including the boundary), respectively.
As usual, we work with the dimensionless variables
\begin{equation}\label{Kandh1}
K=J/k_BT\quad \mbox{and}\quad  h_1=H_1/k_BT\>.
\end{equation}
The bulk critical point corresponds to $K_c=\frac12\, \mbox{ln}
(1+\sqrt{2})$.

\subsection{Scaling analysis for Ising system in $2<d<4$}\label{twoone}

In the critical regime, where $|\tau|\equiv |(T-T_c)/T_c|\ll 1$,
thermodynamic quantities are described
by homogeneous functions of the scaling fields.
As a consequence,
the behavior of the local magnetization under
rescaling of distances should be described by
\begin{equation}\label{scal}
m(z,\tau,{h}_1)\sim b^{-x_{\phi}}\,m(zb^{-1},\,\tau b^{1/\nu},\,
{h}_1\,b^{y_1}),
\end{equation}
where $x_{\phi}=\beta/\nu$ and
$y_1=\Delta_1/\nu$ is the scaling dimension of $h_1$ \cite{diehl}.
In terms of other surface exponents we have
$y_1=(d-\eta_{\parallel})/2$\cite{diehl,foot}.
In Eq.\,(\ref{scal}) it was further assumed
that the distance $z$ from the surface is much larger than
the lattice spacing or any other {\it microscopic} length scale.
One is interested in the behavior at {\it macroscopic} scales, and,
for the present,
$z$ may be considered as a continuous variable ranging
from zero to infinity.

Removing the arbitrary rescaling parameter $b$ in
Eq.\,(\ref{scal}) by setting it
$\sim z$, one obtains the scaling form of the magnetization
\begin{mathletters}\label{scal2}
\begin{equation}\label{scalm}
m(z,\tau,h_1)\sim z^{-x_{\phi}}\,{\cal M}(z/\xi, z/l_1)\>,
\end{equation}
where, as already stated,
\begin{equation}\label{length}
l_1\sim h_1^{-1/y_1}
\end{equation}
\end{mathletters}
\noindent
is the length scale set by the surface field. The second length
scale pertinent to the semi-infinite system and occurring in
(\ref{scalm}) is the bulk correlation length $\xi=\tau^{-\nu}$.
Regarding the interpretation of MC data, which are normally obtained
from finite lattices, one has to take into account a third
length scale, the characteristic dimension $L$ of the system,
and a finite-size scaling analysis
has to be performed. The latter will be briefly described
in Sec.\,\ref{twothree}.

Going back to the semi-infinite case and setting
$\tau=0$,
the only remaining length scale is $l_1$, and
the order-parameter profile can be written in the
critical-point scaling form
\begin{equation}\label{h1}
m(z,{h}_1)\sim z^{-x_{\phi}}\,{\cal M}_c(z/l_1)\>.
\end{equation}
As said above, for $z\to \infty$ the magnetization decays as $\sim
z^{-x_{\phi}}$
and, thus, ${\cal M}_c(\zeta)$ should approach a constant for
$\zeta\to \infty$.
In order to work out the {\it short-distance}
behavior of the scaling function ${\cal M}_c(\zeta)$,
we demand that $m(z)\sim m_1$ as $z\to 0$.
This means that in general, in terms of macroscopic quantities,
the boundary value of $m(z)$ is {\it not} $m_1$. If
the $z$-dependence of $m(z)$ is described
by a power law, it cannot approach any value
different from zero or infinity as $z$ goes to zero.
However, the somewhat weaker relation symbolized by ``$\sim$'' should hold,
stating that the respective
quantity asymptotically (up to constants) ``behaves as" or
``is proportional to".
This is in accord with and actually motivated by the
field-theoretic short-distance expansion \cite{syma,diehl}, where
operators near a boundary are represented in terms of
boundary operators multiplied by $c$-number functions.

In the case of the three-dimensional Ising model the foregoing
discussion leads to the conclusion
that $m(z)\sim h_1$ because the ``ordinary''
surface---the universality class to which also a free surface
belongs---is paramagnetic
and responds linearly to a small magnetic field\cite{bray}.
The consequence for the scaling function in (\ref{h1})
is that ${\cal M}_c(\zeta)\sim \zeta^{y_1}$, and, inserting
this in (\ref{h1}),
we obtain that the exponent governing the short-distance
behavior of $m(z)$ is given by the difference between
$y_1$ and $x_{\phi}$ (as already stated in
Eq.\,(\ref{power})).
Using the scaling relation $\eta_{\perp}=(\eta+\eta_{\parallel})/2$
\cite{diehl} among anomalous dimensions, one can reexpress
the exponent $\kappa$ as \cite{foot}
\begin{equation}\label{kappa}
\kappa=1-\eta_{\perp}\>.
\end{equation}
In the mean field approximation the value of $\kappa$ is zero, and
one really has $m(z\to 0)=m_1$.
However, a positive value is obtained when fluctuations are taken into account
below the upper critical dimensionality $d^*$.
For instance, the result for the $n$-vector model with $n=1$
(belonging to the Ising universality class)
in the framework of the $\epsilon$-expansion is
$\kappa=\epsilon/6$ \cite{czeri}. Thus, the
magnetization indeed grows as $z$ increases away from the surface.

For the 2-$d$ Ising model the exponent $\eta_{\perp}$
is known {\it exactly} \cite{cardy} and one obtains
$\kappa=3/8$. However, as will be discussed in Sec.\,\ref{twotwo},
the pure
power law found in Ref.\,\cite{czeri} for $d=3$ is modified in two
dimensions by a {\it logarithmic} term, and the exponent $3/8$ cannot
directly be seen in the profile.

The above phenomenological analysis is straightforwardly extended
to the case $\tau>0$. In $d>2$, we may assume that the
behavior near the surface for $z<<\xi$ is unchanged compared
to (\ref{power}), and, thus,
the increasing profiles are also expected slightly above the
critical temperature. The behavior farther away from the surface depends
on the ratio $l_1/\xi$.
In the case of $l_1>\xi$ a crossover
to an exponential decay will take place for $z\simeq \xi$
and the regime of nonlinear decay does not occur.
For $l_1< \xi$ a crossover
to the power-law decay $\sim z^{-\beta/\nu}$ takes place
and finally at $z\simeq \xi$ the exponential behavior sets in.

Below the critical temperature, the short-distance
behavior of the order parameter is also described by a power
law, this time governed by a different exponent, however. The essential
point is that below $T_c$ the surface orders even for $h_1=0$.
Hence, in the scaling analysis the scaling dimension of $h_1$
has to be replaced by the scaling dimension of $m_1$, the
conjugate density to $h_1$,
given by $x_1=\beta_1/\nu$\cite{diehl}. The exponent that describes
the increase of the profile is thus
$x_1-x_{\phi}$ \cite{gompper}, a number that even in
mean-field theory is different from zero ($=1/2$).

Phenomena to some extent analogous to the ones discussed above were
reported for the crossover between {\it special} and
normal transition\cite{brezin,ciach}.
Also near the special transition the
surface field $h_1$ gives rise to a length scale. However, the
respective exponent, the analogy to $\kappa$ in (\ref{power}),
is negative, and, thus, one
finds a profile that monotonously decays for all
(macroscopic) $z$, with different power laws
in the short-distance and the long-distance regime and
a crossover at distances comparable to the length scale
set by $h_1$. However,
{\it non-monotonous} behavior in the crossover
region is a common feature in the case
of the energy density in $d=2$ \cite{mifi} and as well as in
higher dimensionality\cite{eisenriegler}.

The {\it spatial} variation of the
magnetization discussed so far
strongly resembles the {\it time} dependence of the
magnetization in relaxational processes at the critical point.
If a system with nonconserved order parameter (model A) is quenched from a
high-temperature initial state to the critical point, with a small
initial magnetization $m^{(i)}$, the order parameter behaves as $m \sim
m^{(i)}\,t^{\theta}$ \cite{jans}, where the short-time exponent $\theta$
is governed
by the difference between the scaling dimensions of initial
and equilibrium magnetization divided by the
dynamic (equilibrium) exponent\cite{own}. Like the exponent $\kappa$
in (\ref{power}), the exponent
$\theta$ vanishes in MF theory, but becomes positive
below $d^*$.

\subsection{Scaling analysis in $d=2$}\label{twotwo}
During the years,
initiated by the work of McCoy and Wu\cite{mccoy,other},
the 2-$d$ Ising model with a surface magnetic field
received a great deal of attention,
because many aspects can be treated
exactly and it is a simple special
version of the 2-$d$ Ising model in a inhomogeneous bulk
field, a problem to which an exact
solution would be highly desirable.
Some of these exact results, namely those considering the
vicinity of the critical point \cite{fishau,bariev},
will be used in the following as a guiding
line for our phenomenological scaling analysis and
to compare numerical data with.

The dependence of $m_1$ on the surface magnetic field (bulk field $h=0$)
in the 2-$d$ Ising model was calculated exactly by Au-Yang
and Fisher \cite{fishau} in a $n\times \infty$ (strip) geometry.
The limit $n\to \infty$, yielding results for the semi-infinite
geometry, was also considered. Whereas above two dimensions
at the ordinary transition the surface, in a sense, is
paramagnetic, i.e., the response
of $m_1$ to a small $h_1$ is linear, in two dimensions
the function $m_1(h_1)$ has a more complicated form. As summarized
in Appendix A (see Eq.\,(\ref{m1h1})),
there is a logarithmic correction to the
linear term in $d=2$; for $h_1\to 0$ the surface magnetization
behaves as $\sim h_1\,\mbox{ln}\,h_1$.

Further, as shown by Bariev \cite{bariev} and summarized in Appendix B,
the length scale $l_1$ determined by $h_1$ behaves as
$\sim \left[\tanh (h_1)\right]^{-2}$.
For small $h_1$, where the scaling analysis is expected to be
correct,
this is consistent with (\ref{length}) as
$y_1=1/2$ in the 2-$d$ Ising model. Thus the characteristic length
scale that enters the scaling analysis
depends in the same way upon $h_1$ as in higher dimensions.

The foregoing discussion allows us to generalize our scaling analysis,
especially Eqs.\,(\ref{h1}) and the near-surface law (\ref{power}),
such that the speacial features of the 2-$d$ Ising model are taken into
account.
Again, the only available length scale at
$\tau=0$ is $l_1$, and
the magnetization can be represented in the form given
in Eq.\,(\ref{h1}).
For $z\to \infty$ we expect that $m\sim z^{-1/8}$
and, thus, ${\cal M}_c(\zeta)$ should approach a constant for
$\zeta\to \infty$.
In order to find the short-distance behavior
we assume again that $m(z)\sim m_1$ as $z\to 0$.
Taking into account the logarithmic correction
mentioned above and discussed in more detail in Appendix A
(see Eq.\,(\ref{m1h1}), we find that
${\cal M}_c(\zeta)\sim \zeta^{1/2}\,\mbox{ln}\,\zeta$
for $\zeta\to 0$. Hence, for the short-distance behavior of $m(z)$
in the semi-infinite system we obtain
\begin{equation}\label{sdbe}
m(z,h_1)\sim h_1\, z^{\kappa}\,\mbox{ln}(h_1\,z^{y_1})\>,
\end{equation}
where the exact values of the exponents are
$\kappa=1-\eta_{\perp}=3/8$ and $y_1=1/2$.
Thus, for $z < l_1$ the magnetization $m(z)$
for a given value of $h_1$ behaves as $\sim z^{3/8}\,\mbox{ln}\,z$.

The result (\ref{sdbe}) should hold for any value of the exchange
coupling $J_1$ in the surface. In our MC analyses to be presented below
we implemented free boundary conditions with
$J=J_1$, but we expect (\ref{sdbe}) to hold for any value of $J_1$ with
possible $J_1$-dependent nonuniversal
constants leaving the qualitative behavior of the profiles unchanged.

Eq.\,(\ref{sdbe}) is the main analytic result of this work.
As discussed in the following,
it is consistent with Bariev's exact solution\cite{bariev}
(see Appendix B) and with MC data for the profile.
It tells us that the short-distance power law behavior is
modified by a logarithmic term. This logarithm
can be traced back to the logarithmic singularity
of the surface
susceptibility\cite{binder}, which, in turn,
causes the logarithmic dependence
of $m_1$ on $h_1$, and eventually leaves its fingerprint
also on the near-surface behavior of the magnetization.
The result (\ref{sdbe}) provides
a thorough understanding of the near-surface
behavior of the order parameter and allows to relate special
features of the two-dimensional system, which were (as
we will discuss in more detail below)
previously known from the exact analyses \cite{bariev}, to the
somewhat simpler short-distance law in higher dimensions.

\subsection{Finite Size scaling}\label{twothree}
In order to assess the finite size effects to be
expected in the MC simulations, we have to take into account the
finite-size length scale $L$, which is proportional to the linear
extension $N$ of the lattice (compare Sec.\,\ref{threeone} below).
The generalization of (\ref{scal}) reads\cite{fisi}
\begin{equation}\label{scalfs}
m(z,\tau,{\sf h}_1,L)\sim b^{-x_{\phi}}\,m(zb^{-1},\,\tau b^{1/\nu},\,
{h}_1\,b^{y_1}, Lb^{-1})\>,
\end{equation}
and proceeding as before, we obtain as the generalization
of (\ref{scalm}) to a system of finite size:
\begin{equation}\label{scalmfs}
m(z,\tau,h_1,L)\sim z^{-x_{\phi}}\,{\cal M}(z/\xi, z/l_1,z/L)\>.
\end{equation}
Thus even at $T_c$ there are two pertinent length scales,
on the one hand $L$ (imposed by
the geometry that limits the wavelength of fluctuations)
and on the other hand $l_1$ (the scale set by $h_1$).

It is well known that for large $z\gtrsim L$ we have to expect
an exponential decay of $m(z)$ on the
scale $L$. In the opposite limit,
when $z$ is smaller than both $L$ and $l_1$, we expect
the short-distance
behavior (\ref{sdbe}) to occur.
However, for $d=2$ it can be concluded from the
finite-size result (\ref{m1h1fin})
that there will be an $L$-dependent amplitude, a prefactor
to the function given in (\ref{sdbe}). But otherwise
the logarithmically modified power law (\ref{sdbe}) should occur.
Farther away from the surface, the form of the profile
depends on the ratio between $l_1$ and $L$.
For $z$ smaller than both $L$ and $l_1$, the behavior
is described
by (\ref{sdbe}). In the case of $l_1>L$ a crossover
to an exponential decay will take place for $z\simeq L$. In the
opposite case, a crossover
to the power-law decay $\sim z^{-x_{\phi}}$ takes place,
followed by the crossover to the exponential
behavior at $z\simeq L$. Thus, qualitatively,
the discussion is
completely analogous to the one in Sec.\,\ref{twoone},
where the behavior at finite $\xi$ was described.

\section{Monte Carlo simulation}\label{three}
\subsection{Method}\label{threeone}
The results of the scaling analysis, especially the
short-distance law (\ref{sdbe}), were checked by MC simulations.
To this end, we calculated order-parameter profiles for the 2-$d$ Ising model
with uniform exchange coupling $J$.
The geometry of our systems was that of
a rectangular (square)
lattice with two free boundaries (opposite to each other)
and the other boundaries periodically coupled, such that the
effective geometry was that of a cylinder of finite length.
The linear dimension perpendicular to the free surfaces
was taken to be four times larger than the lateral extension
in order to keep corrections due to the second surface
small\cite{fidege}. Hence,
when we talk about a lattice of size $N$ in the
following, we refer to a rectangular
$N\times 4\,N$ system.

In order to generate an equilibrium sample of spin
configurations, we used
the Swendson-Wang algorithm\cite{swewa}.
It effectively avoids critical slowing
down by generating new spin configurations
via clusters of bonds, whereby
the law of detailed balance is obeyed.
For a given spin configuration, a bond between two neighboring
spins of {\it equal} sign exits with probability $1-e^{-2K}$.
There are no bonds between {\it opposite} spins.
Then, clusters are defined
as any connected configuration of bonds.
Also isolated spins define
a cluster, such that eventually
each spin belongs to one of the clusters.
After having identified the clusters,
the new configuration is generated by assigning to
each cluster of spins a new orientation, with equal
probability for each spin value as long as the cluster does {\it not}
extend to a surface.

In order to take into account $h_1$, we introduced,
in the same way as
suggested by Wang for taking into account bulk fields
\cite{wang},
two ``ghost'' layers of spins next to
each surface that couple to the surface spins with
coupling strength $h_1$ and that all point in the direction of $h_1$.
If at least one bond between
a surface and a ghost spin exists the
cluster has to keep its old spin
when the system is updated.  This preserves
detailed balance.
In the practical calculation this
rule was realized by a modified (reduced) spin-flip probability
\begin{equation}
p(k)=1-\frac{1}{2} \,\exp(-2\,h_1\, k)
\end{equation}
for clusters pointing in the direction of $h_1$ (and $1/2$
for clusters pointing in opposite direction),
$k$ being the number of {\it surface}
spins contained in the cluster.

In order to obtain an equilibrium distribution of configurations
we discharged several hundred (depending on system size)
configurations after the start of the simulation.
To keep memory consume low, we used multispin-coding techniques,
i.e., groups of 64 spins were coded in one long integer.

\subsection{Comparison with exact results}\label{threetwo}
A crucial test for the MC program is the comparison with
known exact results. On the other hand, if both MC data
an exact results agree, the former can be regarded also
as a confirmation of the exact results.
Having calculated order-parameter profiles
for different values of $h_1$, in particular
the magnetization at the boundaries
(the ends of the cylinder) can be obtained.
In Fig.\,1 we show the results for $m_1(h_1)$. The squares
respresent our MC data for a system of size $N=512$.
Also depicted is the exact result for the semi-infinite system
taken from Ref. \cite{fishau} (see
also Appendix\,\ref{appa}). It is clear that the MC
values approach the exact curve for large $h_1$.
Below $h_1\simeq 0.03$,
the MC results show a linear dependence on $h_1$,
significantly deviating from the exact curve.
The linear behavior can be regarded as a finite-size effect,
and it is qualitatively consistent with the result (\ref{m1h1fin})
of Au-Yang and Fisher\cite{fishau}. The latter was
derived in a strip of finite width with infinite lateral extension,
however,
so that a quantitative comparison with our data is not possible.\\[2mm]

\def\epsfsize#1#2{0.6#1}
\hspace*{2.5cm}\epsfbox{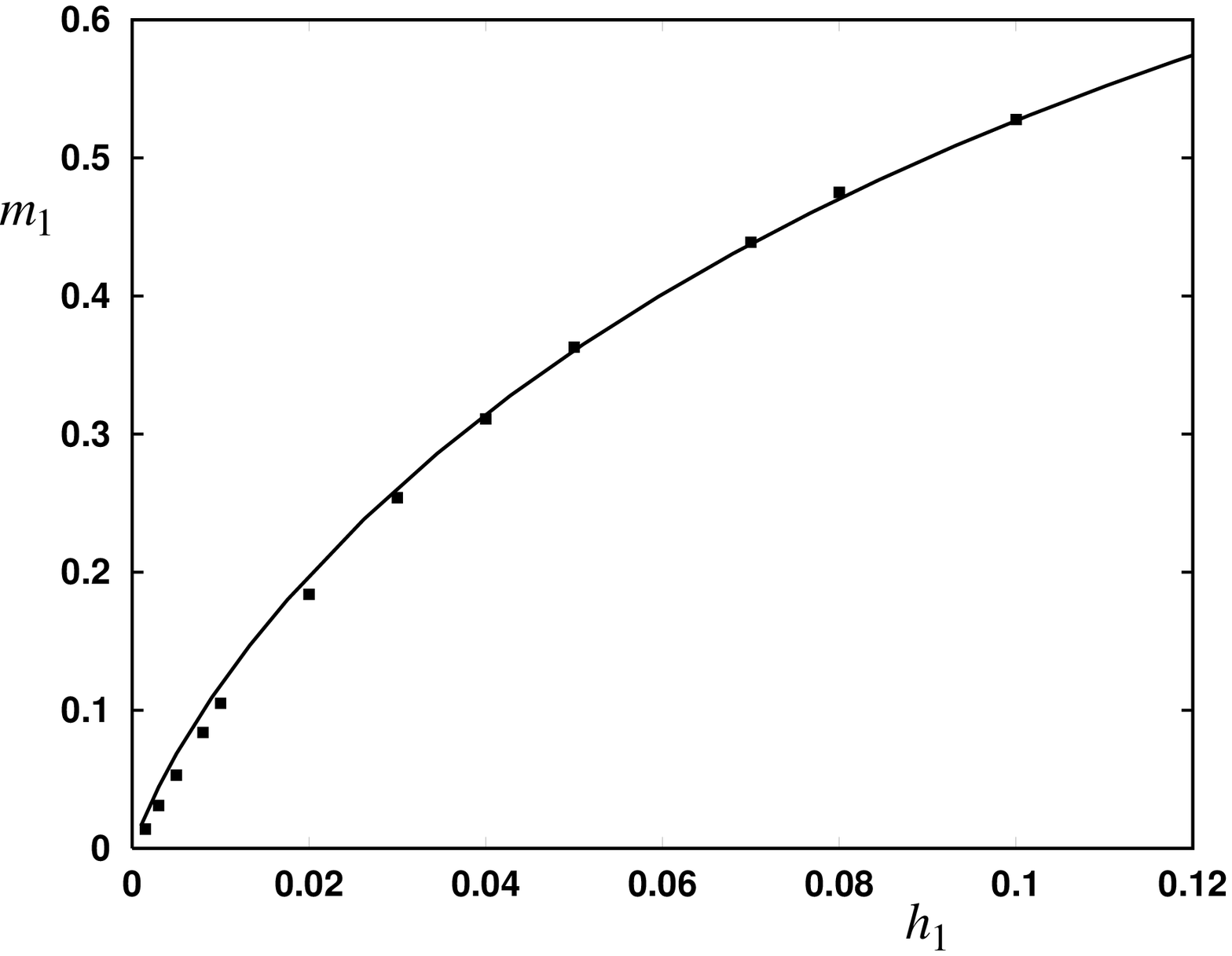}\\[0mm]
{\small {\bf Fig.\,1}: Monte Carlo results for $m_1$ as a function of $h_1$
for $N=512$ (represented by full squares)
compared to the exact result of
Ref.\,\cite{fishau} (see Appendix A). The statistical errors of the
Monte Carlo data are about the same size as the symbols. A detailed
discussion and comparison of the data is in the text.} \\[0.1cm]

Next we compare with the exact solution obtained
by Bariev \cite{bariev} for the order-parameter profile.
The explicit result (\ref{asympt}) holds in the limit
$h_1\to 0$. This limit is hard to access in the MC simulation
since the signal $m(z)$ becomes small and is
eventually lost in the noise.
To obtain a concrete result to compare with, we have calculated
the profile numerically from (\ref{bariev}) terminating the
series in (\ref{exact2})
after the third term. It turned out that
the series converges rather quickly as long as
the distance from the surface is not too small.
Only very close to the surface
higher orders need to be taken into account.
Concretely, we took $K=0.999 K_c$ (i.e. $\tau \simeq 0.001$) and $h_1=0.01$.
Then, employing (\ref{correl}) and
(\ref{bariev}) one obtains
$\xi=567$ and $l_1=2069$, respectively. The result
of the numerical evaluation of (\ref{bariev}) is shown in Fig.\,2
(dashed curve), where $m(z)$ versus the distance $z\equiv n-1$
is plotted.

Then, with exactly
the same parameters, the MC profiles were calculated,
with the size $N$ varying between 128 and
1024. The results are depicted in double-logarithmic representation
in Fig.\,2 (solid curves). It is obvious that the MC data approach
the (approximated, in principle) exact profile
with increasing lattice size. Most importantly, both
results show the short-distance behavior anticipated from the
scaling analysis above and expressed in (\ref{sdbe}).
This is demonstrated by the dotted line, which depicts
the function $0.016\, n^{3/8}\,(6.15-\mbox{ln}\,n)$ (the
constants were fitted). As expected, it only describes
the profile for short distances from the boundary and becomes
wrong for large distances, completely analogous to the asymptotic
form (\ref{m1h1}) of $m_1(h_1)$.
\\[2mm]

\def\epsfsize#1#2{0.6#1}
\hspace*{2.5cm}\epsfbox{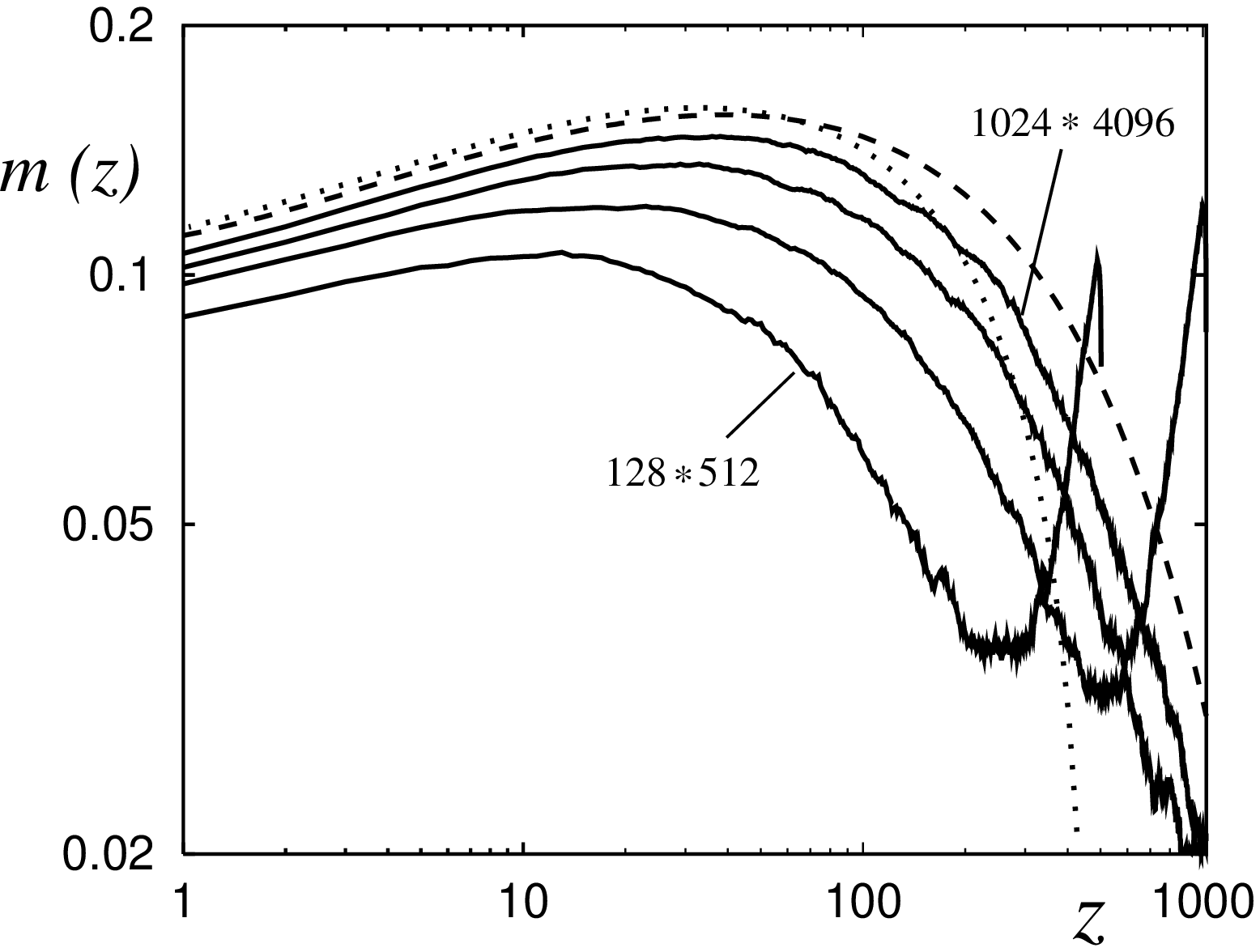}\\[0mm]
{\small {\bf Fig.\,2}: Monte Carlo profiles for $K/K_c=0.001$ and $h_1=0.01$
for lattices of size 128$\times$512, 256$\times$1024, 512$\times$2048, and 1024$\times$4096 (solid lines from below to above) compared with
the numerical evaluation of the exact result (\ref{bariev}). The latter
holds for the semi-infinite system. It is clearly visible that the
Monte Carlo data approach the exact result for increasibng system
size. The dotted line represents the
asymptotic ($z\to 0$)
behavior expressed in Eq.\,(\ref{sdbe}) $\sim z\,\mbox{ln}\,z$
that becomes wrong for
large distances. }\\[-.8cm]

\subsection{Monte Carlo results}
\label{threethree}
First we discuss a set of profiles which were obtained with
$N=512$
by setting $h_1=0.01$ and $K/K_c$ varying between 0.996 to
1.004, in steps of 0.001. The data are depicted in Fig.\,3.
Depending on the temperature, we averaged over 10\,000 to
30\,000 configurations. Especially the shape
of the critical profile, marked
in Fig.\,3, is consistent with
the scaling analysis of Sec.\,\ref{twotwo}.
It increases up to $z\simeq 60$, then has a maximum
and farther away from the surface it decays.
With increasing distance the influence of the second surface
(here at $z=2047$) becomes stronger, such that the profile
has a minimum about halfway between the boundaries. (The data
of Fig.\,3 were not symmetrized after the average over configurations
was taken.)

For $T$ above $T_c$ (curves below the critical profile), the
maximum moves towards the surface and the regime of growing
magnetization becomes
smaller. This is consistent with the scaling analysis
of Sec.\,\ref{twoone}, in this case the growth is limited
by the correlation length $\xi$.
On the other hand, for $T<T_c$, the tendency to decay
in between the surfaces becomes weaker, as the the bulk
value of the magnetization in the ordered phase grows.
As said in Sec.\,\ref{twoone}, the short-distance
growth below $T_c$ is described by $z^{x_1-x_{\phi}}$, where the
difference $x_1-x_{\phi}$ takes also the exact
value 3/8 \cite{foot5}. This time, there is no logarithm
however, and the growth is steeper than above $T_c$.
\\[2mm]

\def\epsfsize#1#2{0.6#1}
\hspace*{2.5cm}\epsfbox{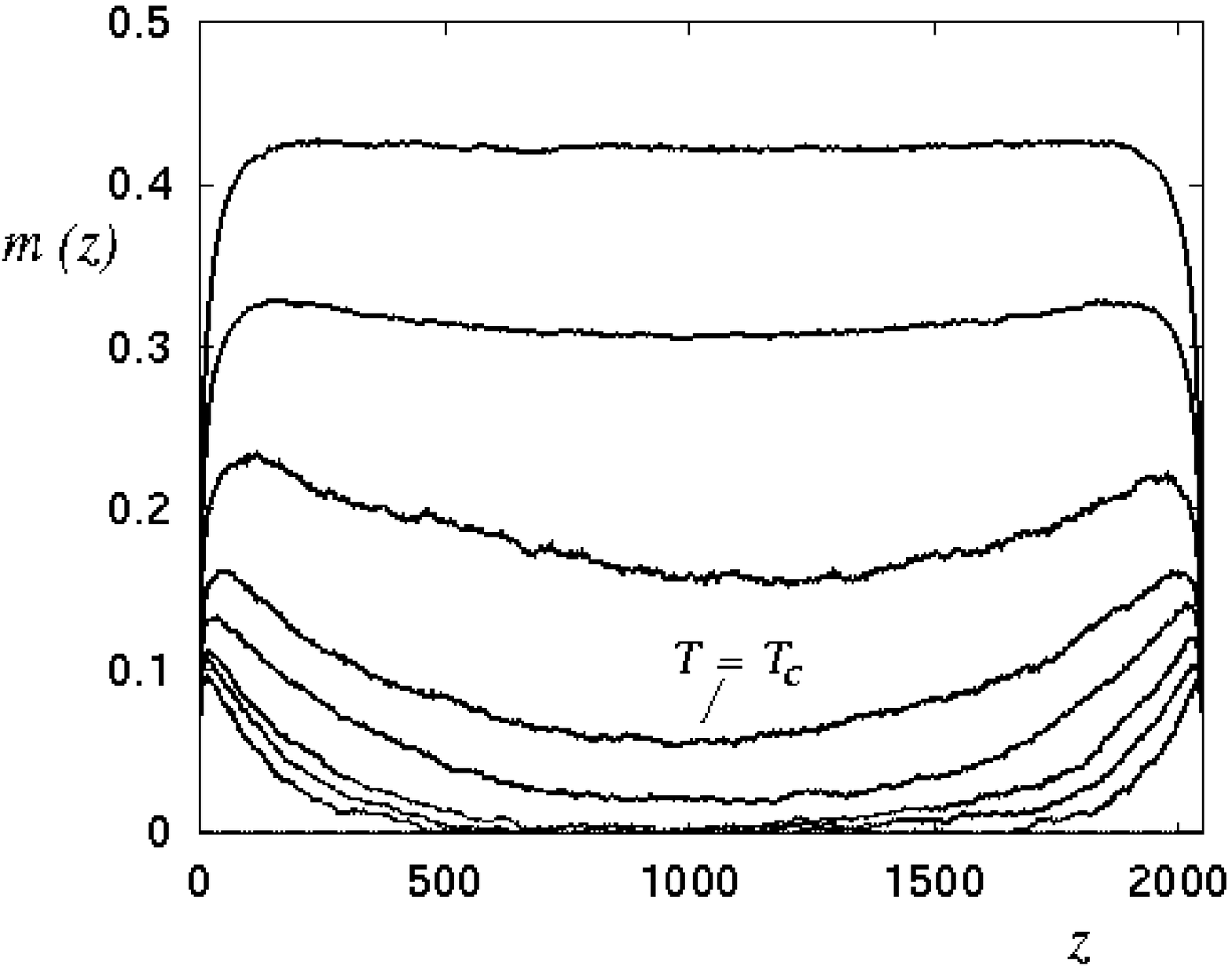}\\[0mm]
{\small {\bf Fig.\,3}: Order-parameter profiles for various temperatures
below and above $T_c$ for fixed $h_1=0.01$ and $N=512$
compared with the critical profile.
The three curves
above the critical profile
were obtained with $K/K_c = 1.001$, 1.002, 1.003.
The four profiles below correspond to
$K/K_c=0.999 \ldots 0.996$. The data are not symmetrized.}\\

Fig.\,4 shows the MC results for the critical point, for
different values of $h_1$ in double-linear representation.
In Fig.\,5 the same data are plotted double-logarithmically.
The lower dashed line shows the short-distance behavior $\sim z^{3/8}\,
\mbox{ln}\,z$ according to
Eq.\,(\ref{sdbe}) and (as already discussed in connection with
Fig.\,2) the MC profiles
confirm the scaling analysis of Sec.\,\ref{twotwo}.
The upper dashed line is the pure power law $z^{-1/8}$ that
describes the decay in the regime where $z$ is
larger than $l_1$ but still smaller
than $L$ or $\xi$. In our simulation this regime is only reached for
relatively large values of $h_1\gtrsim 0.5$.  The uppermost profile ($h_1=1.0$)
obviously goes through
this regime for $10\lesssim z\lesssim 60$. The finite-size
exponential behavior (see Sec.\ref{twothree}) can be observed in all
curves for $z\gtrsim 100$. The data of Figs.\,4 and 5 are symmetrized, i.e.,
after averaging over configurations we computed the mean value
of the left and right halfs of the system. Hence, the profiles
are only displayed up to halfway between the
boundaries (here $z=1023$).

The location of the maximum of $m(z)$,
$z_{\rm max}$, as a function of $h_1$
is depicted in Fig.\,6. The maximum $z_{\rm max}$ was determined from the
profiles by a graphical method. Error bars are estimated.
For small values of $h_1$ the near-surface growth
is limited by finite-size effects.
Up to about $h_1=0.03$, the value of
$z_{\rm max}$
is roughly independent of $h_1$.
For larger values of $h_1$, $z_{\rm max}$
moves towards the surface. As indicated by the dashed line, the
dependence of $z_{\rm max}$ on $h_1$ is completely consistent with
$l_1 \sim h_1^{-2}$ obtained from the scaling analysis (see (\ref{length})).
For $h_1=1.0$ (upper curve) the maximum value
of $m(z)$ is at the boundary, and the
magnetization monotonously decays for $z>0$.\\[2mm]

\def\epsfsize#1#2{0.6#1}
\hspace*{2.5cm}\epsfbox{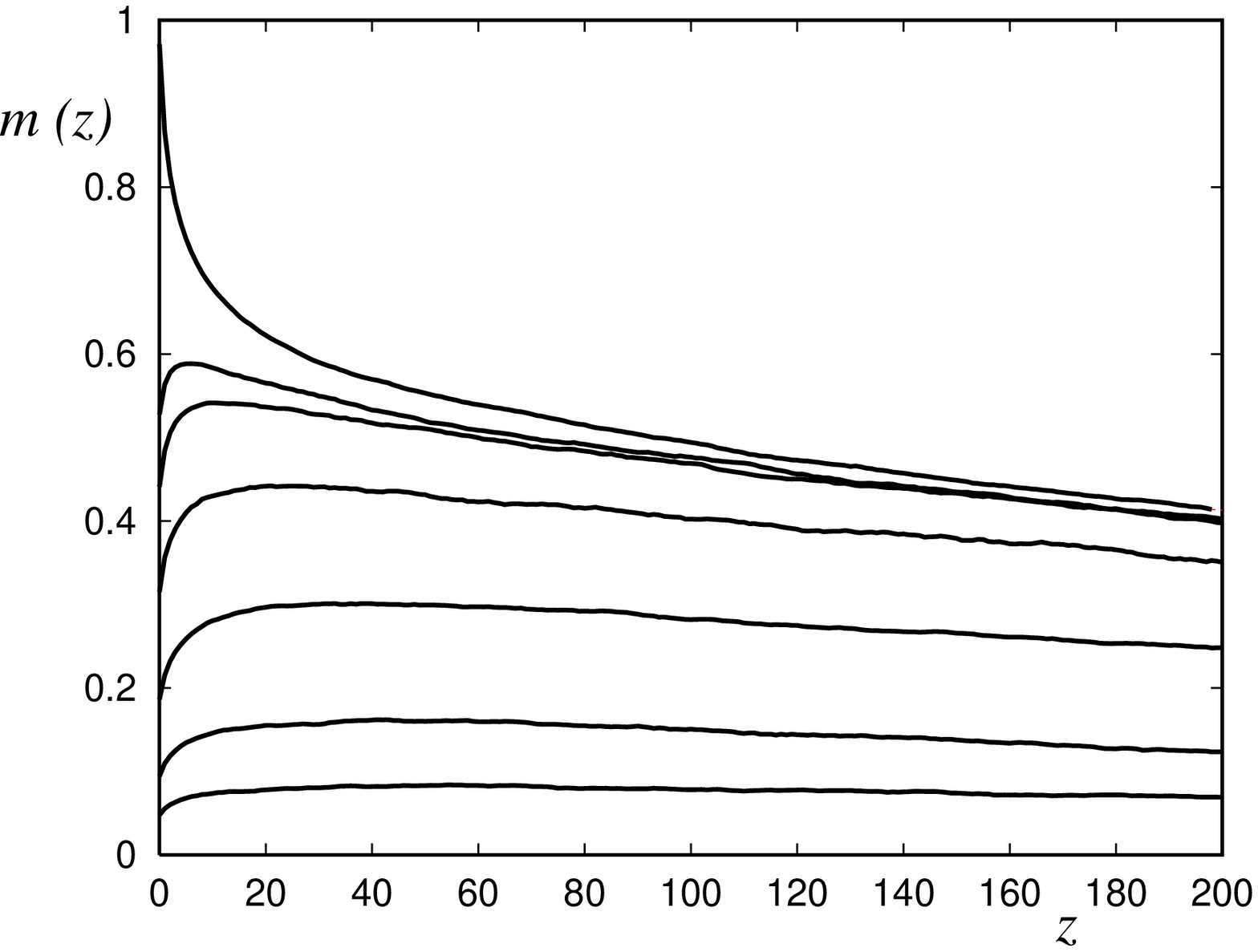}\\[0mm]
{\small {\bf Fig.\,4}: Monte Carlo profiles for $N=512$ at $T=T_c$ for
$h_1=0.005$, 0.01, 0.02, 0.03, 0.07, 0.1, and 1.0
(from bottom to top).
The data are symmetrized and are
only shown up to $z=1023$.} \\[0.1cm]

\def\epsfsize#1#2{0.6#1}
\hspace*{2.5cm}\epsfbox{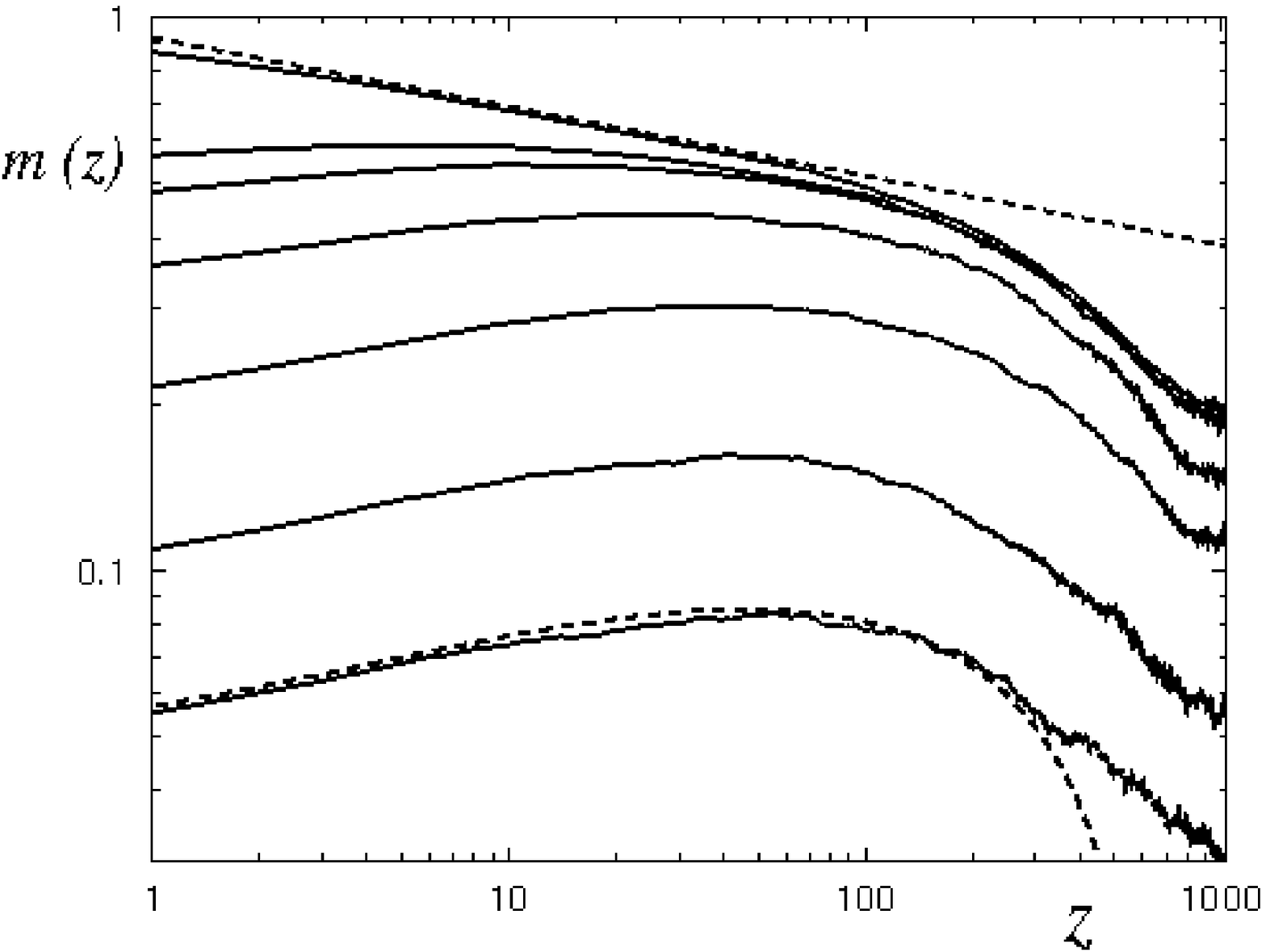}\\[0mm]
{\small {\bf Fig.\,5}: The same profiles as in Fig.\,4 in double-logarithmic
representation, pronouncing the short-distance behavior.
For small $h_1$ (lower curves) the growth of $m(z)$ described
by Eq.\,(\ref{sdbe}) is clearly visible
For $h_1=1.0$ our Monte Carlo
result (upper solid line) is in accord with $m(z)\sim z^{-1/8}$
(dashed line). The behavior of the Monte Carlo data for $z\gtrsim 100$
is described by the (finite-size) exponential decay and corrections
to the semi-infinite profiles due to the second boundary become stronger.} \\[0.1cm]

\def\epsfsize#1#2{0.6#1}
\hspace*{2.5cm}\epsfbox{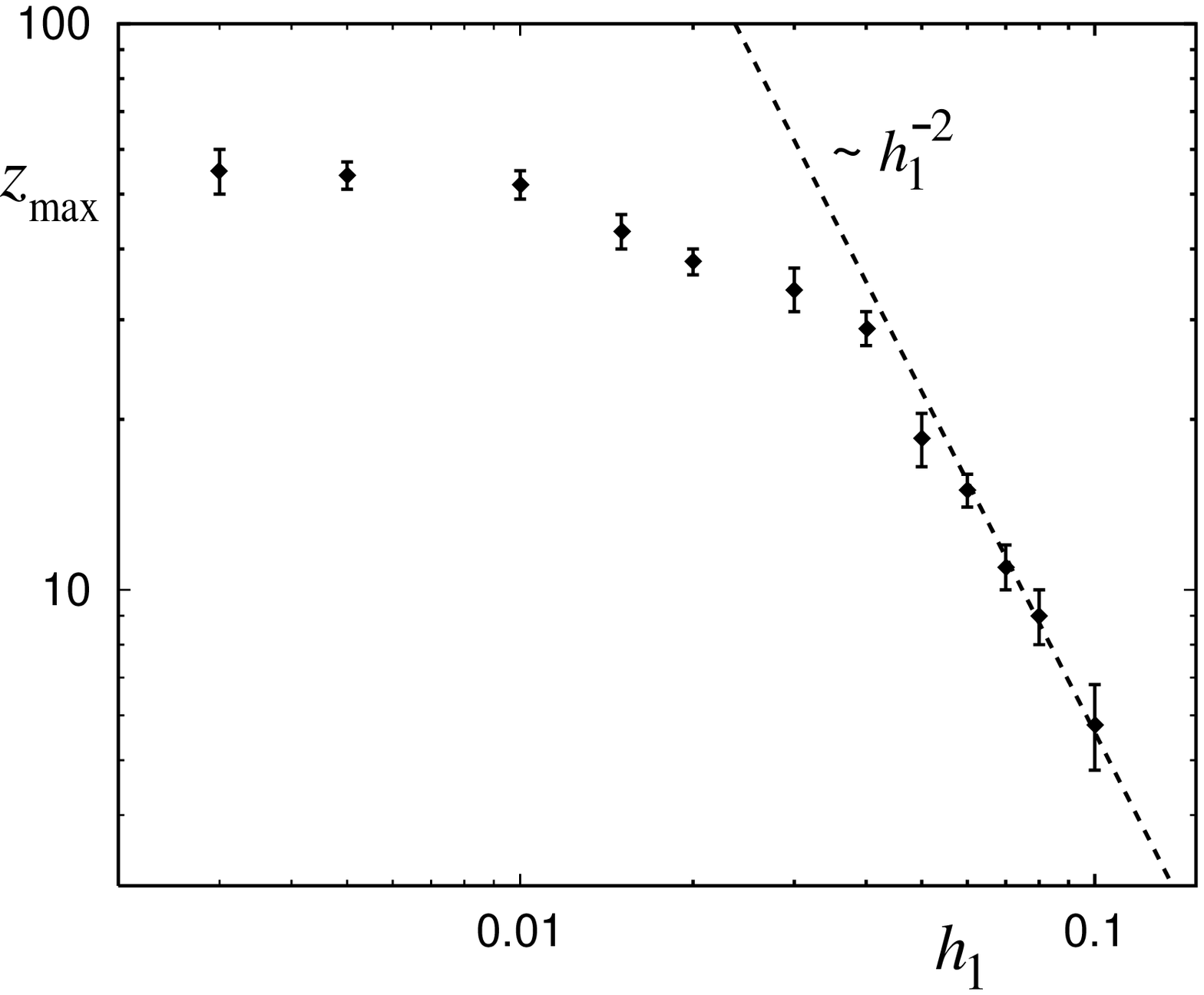}\\[0mm]
{\small {\bf Fig.\,6}: The location of the maximum $z_{\rm max}$ of $m(z)$
in dependence of $h_1$ as obtained from the data of Fig.\,4 and
other profiles for the same system
(not displayed). The dashed line represents
$l_1\sim h_1^{-2}$. For larger values of $h_1$ we have
$z_{\rm max}\sim l_1$. For small $h_1$ where $l_1$
becomes larger than the finite-size scale $L$, $z_{\rm max}$
is determined by $L$.} \\[0.1cm]

\def\epsfsize#1#2{0.6#1}
\hspace*{2.5cm}\epsfbox{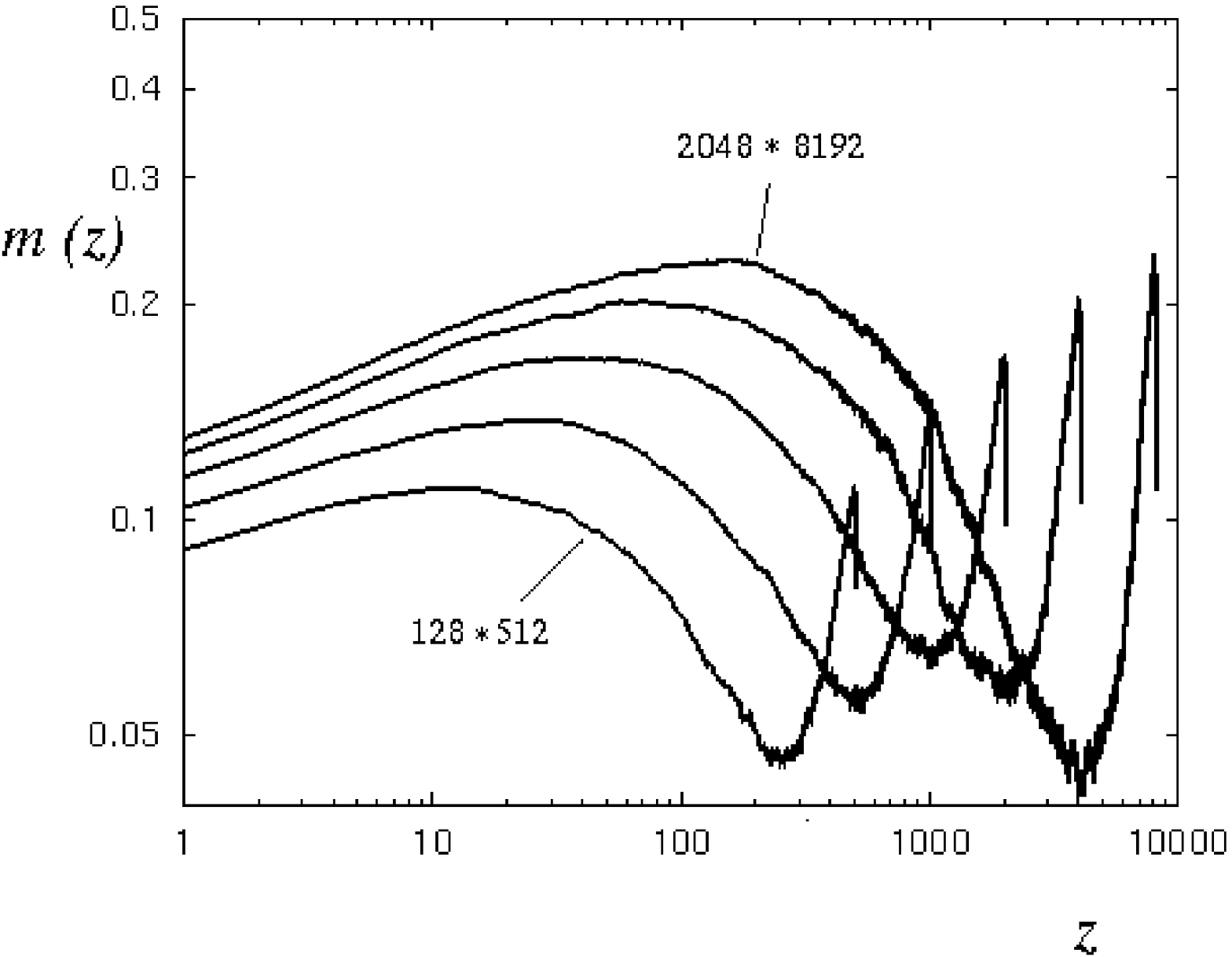}\\[0mm]
{\small {\bf Fig.\,7}: Order-parameter profiles for $K=K_c$ and $h_1=0.01$
for system size $N=128$, 256, 512, 1024, and 2048.
} \\[0.1cm]

In Fig.\,7 profiles for fixed $h_1$ and $T=T_c$ for
various system sizes (ranging between $N=128$ and 2048) are displayed.
With increasing $N$, the maximum keeps moving away from the
surface, $z_{\rm max}\simeq$ in the largest system.
This means that with these parameters we are still in the regime
where $L<l_1$. When $h_1=0.03$ is taken instead, the situation
is different. The respective MC data are depicted in
Fig.\,8. For the small system we have again increasing $z_{\rm max}$.
For the two largest systems $N=1024$ and 2048, however, the
maximum is roughly located at the same
distance from the surface, signaling that now $l_1<L$.\\[3mm]

\def\epsfsize#1#2{0.6#1}
\hspace*{2.5cm}\epsfbox{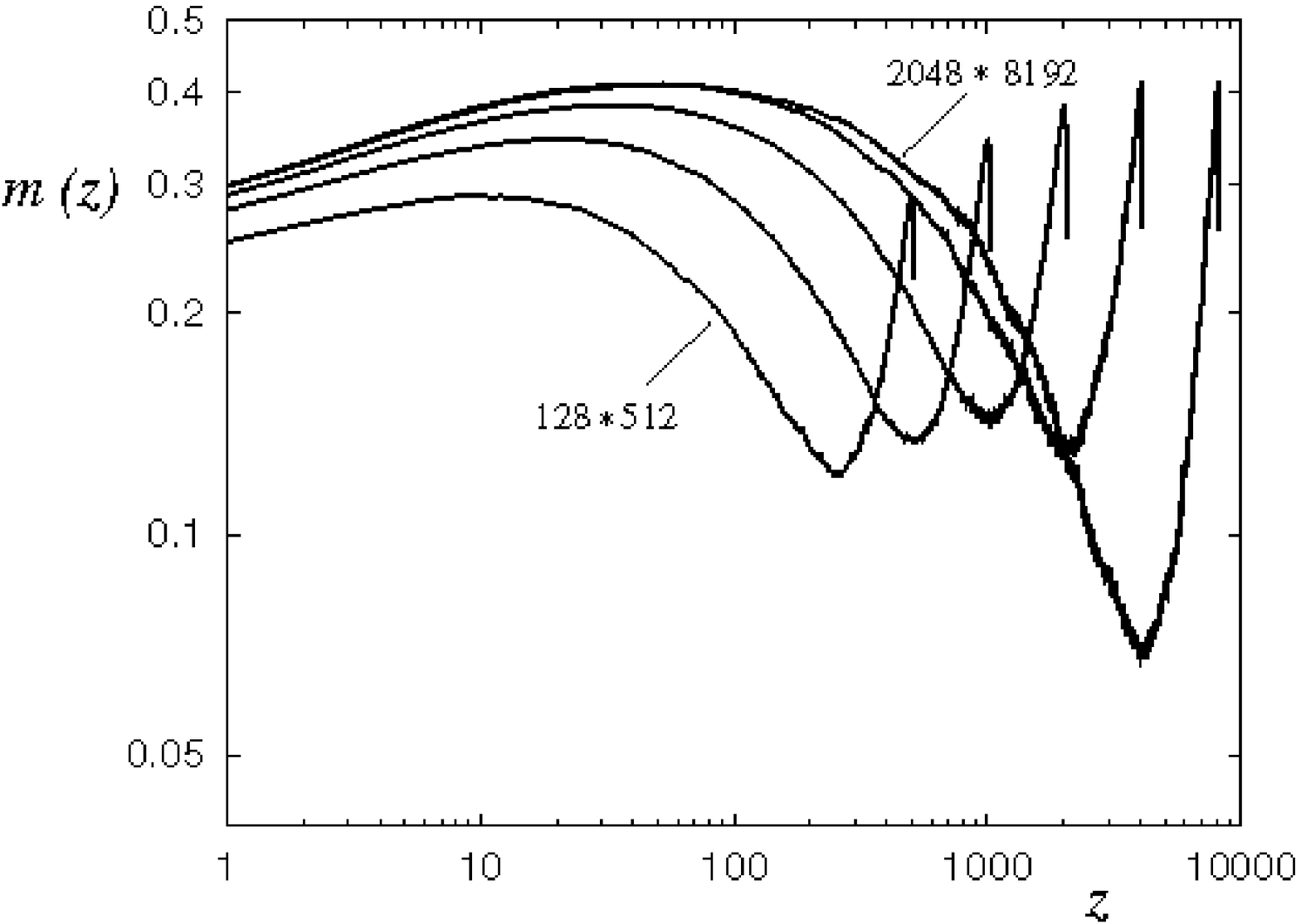}\\[0mm]
{\small {\bf Fig.\,8}:  Profiles for $h_1=0.03$ and otherwise
the same parameters as in Fig.\,7.
} \\[0.1cm]

\section{Discussion} \label{disc}
We studied the short-distance behavior of order-parameter profiles
in the two-dimensional semi-infinite Ising system at or above
the bulk critical temperature. Our main goals
were a detailed understanding of the near-surface
behavior of the order parameter and
a complete scenario for the crossover between
``ordinary'' and ``normal'' transition.

With regard to the short-distance behavior, especially for
small surface fields $h_1$, the Ising model in $d=2$
turned out to be quite special. The
functional form is
quantitatively not captured by
the analysis of Ref.\,\cite{czeri},
where the main emphasis was put on the situation in $d=3$.
Here we demonstrated by means of a scaling analysis
that a small $h_1$ induces a magnetization
in a surface-near regime that grows as $z^{3/8}\,\mbox{ln}\,z$
as the distance $z$ increases away from the surface
(see Eq.\,(\ref{sdbe})).
Our findings are not only consistent
with available exact results (to which we made a detailed comparison),
but they also allow a detailed understanding of the
physical reasons for the growth and its quantitative
form. The surface magnetization $m_1$ generates in the region that is
(on macroscopic scales) close to the surface and that
is much more suceptible than the surface itself, a magnetization
$m(z)$ much larger than $m_1$.
The exponent $3/8$ that governs the power-law part of the growth
is the difference between the scaling dimensions $y_1$ (of $h_1$)
and $x_{\phi}$ (of the bulk magnetization).
Eventually, as was demonstrated Sec.\,\ref{twotwo},
the logarithmic factor can be traced back to the logarithmic
dependence of $m_1$ on $h_1$.

The scenario for the crossover
between ``ordinary'' and ``normal'' developed
in Ref.\,\cite{czeri} and generalized
in this work to include the 2-$d$ Ising model is
the following: At $T=T_c$ a small $h_1$ causes the increasing
near-surface behavior described above. The magnetization
grows up to distances $l_1\sim h_1^{-2}$ and then
the crossover to the power-law decay $\sim z^{-1/8}$ takes place.
With increasing $h_1$ the surface-near regime becomes smaller, and
eventually for $h_1\to\infty$ the length scale $l_1$ goes to zero, such
that the region with increasing magnetization vanishes completely
and the situation of the ``normal'' transition is reached.

For $T$ slightly above $T_c$ our scenario essentially
remains valid as long as
$\xi >l_1$. Only when $z$ is of the order of $\xi$ the
crossover to the exponential decay takes place. For $\xi < l_1$,
on the other hand, the
growth is limited to the region $z<\xi$.

Concerning three-dimensional systems
several experiments were pointed out in Ref.\,\cite{czeri},
whose results are possibly related
related to the anomalous short-distance behavior \cite{mail,franck}.
It would be an interesting question for the future, whether
similar, surface-sensitive
experiments in two-dimensional systems are feasible.

{\small {\it Acknowledgements}: We thank R. Z. Bariev, E. Eisenriegler,
and M. E. Fisher for helpful comments.
This work was supported in part by the Deutsche Forschungsgemeinschaft
through Sonderforschungsbereich 237.}

\appendix

\section{Exact results for the surface magnetization}\label{appa}
The boundary magnetization $m_1(h_1)$ was calculated
exactly by Au-Yang and Fisher \cite{fishau}
for the semi-infinite
2-$d$ Ising model at the critical point, in the
critical regime, and for a strip of finite width at the
bulk critical point.
Whereas above two dimensions a free surface
near criticality is paramagnetic in the sense that
the response of $m_1$ on $h_1$ is linear,
it was found that in $d=2$ the function $m_1(h_1)$ is
modified by logarithmic terms.

In the 2-$d$ Ising model the surface magnetic field $h_1$ enters
the analysis only through the variable \cite{fishau}
\begin{equation}\label{u}
u=\tanh \,h_1\>,
\end{equation}
which will also play an important role in Appendix \ref{appb}.
For $\tau=0$ and small $u$ (in which case $u\approx h_1$) the
asymptotic behavior of $m_1$ takes the form\cite{fishau}
\begin{equation}\label{m1h1}
m_1(h_1)\approx B_1\,u\,(b_1-2\, \mbox{ln}\,u)\>,
\end{equation}
where $B_1=1.5369\ldots$ and $b_1=-1.3201\ldots$ are constants.
There is an
exact formula available for $m_1(h_1)$ (see Eqs.\,(2.10) and
(2.11) of Ref.\,\cite{fishau})
but the asymptotic result (\ref{m1h1})
was sufficient for carrying out the scaling analysis in
Sec.\,\ref{twotwo}.
In Sec.\,\ref{threetwo}, we
compare our MC data for $m_1$ with the exact result.
The lesson learned from (\ref{m1h1}) is that
in $d=2$ and in semi-infinite geometry
the dependence of $m_1$ and $h_1$ is {\it not}
simply linear but modified by a logarithmic term:
For small $h_1$ the surface magnetization behaves as
$h_1\,\mbox{ln}\,h_1$.
This was not taken into account in Ref.\,\cite{czeri},
where we focussed on the short-distance behavior in $d=3$.

In the critical region, for $\tau>0$, the leading
contribution to the surface magnetization is given by\cite{fishau}
\begin{equation}\label{m1tau}
m_1(h_1,\tau)= B_1\,u\,\left[\mbox{ln}(2K_c\tau)+{\cal B}(\tau/u^2)
\right]\>.
\end{equation}
This tells us that
slightly above the critical point the dependence of $m_1$
on $h_1$ is indeed linear, but, different from the situation in
$d>2$, there is a (logarithmic) dependence on the
temperature $\tau$. The scaling function ${\cal B}$ approaches
a constant for $\tau/u^2\to 0$ and asymptotically goes as
$\sim \mbox{ln}(\tau/u^2)$ for $\tau/u^2\to \infty$, leading
back to (\ref{m1h1}).

In Ref.\,\cite{fishau} the function $m_1(h_1)$ was also calculated
for a strip of finite width $N$ and infinite lateral extension at
the (bulk) critical point.
This result is of interest for the comparison with our MC
data. Similarly to the case $\tau>0$, the logarithmic term
in the limit $u\to 0$ is absent in the strip.
The leading contribution can be written in  the form
\begin{equation}\label{m1h1fin}
m_1(h_1)=B_1\,u\,(b_1^*+c_1^*\,N^{-1}+\mbox{ln}\, N)\>,
\end{equation}
where $b_1^*=1.0731\ldots$ and $c_1^*=\sqrt{2}/2$. Thus, as in
(\ref{m1tau}) the dependence on $h_1$ is linear, with
an $N$-dependent amplitude $\sim \mbox{ln}\,N$.
Taking this into account in the scaling analysis, it leads
to an $N$-dependent amplitude in the profile, an effect that
is clearly visible in Figs.\,2, 7 and 8, where the profiles
for different system sizes were plotted.

\section{Exact results for the magnetization profile}\label{appb}
In a work by Bariev \cite{bariev}, the local magnetization
of the semi-infinite 2-$d$ Ising model
with surface magnetic field was represented
in terms of a series of multiple integrals.
For the system we consider \cite{foot4} this solution
can be written in the form
\begin{equation}\label{exact1}
m(n,K,h_1)=\left(1-S^2\right)^{1/8} {\cal F}(2n/\xi,l_1/\xi)\>,
\end{equation}
where $S=\sinh(2K)$, the integer $n$ counts the distance
(number of spin columns) from the boundary
starting with $n=1$, and
$\xi$ is the correlation length defined by
\begin{equation}\label{correl}
\xi=\sqrt{\frac{S}{2}}\,\frac{1}{1-S}\>.
\end{equation}
The dimensionaless variables
$K$ and $h_1$ were defined in (\ref{Kandh1}).

The length scale $l_1$ occuring in (\ref{exact1})
represents the distance
up to which the short-distance behavior
(\ref{sdbe}) is found.
It is explicitely given by \cite{bariev}
\begin{equation}
l_1=\frac{\tanh(K)}{2}\, \frac{1}{u^2}\>.
\end{equation}

The main result of Ref.\,\cite{bariev} is
the scaling function
${\cal F}(x,y)$ in (\ref{exact1})
expressed in form of a series of multiple integrals, with
distinct solutions
above and below the critical temperature.
Since we are interested in the
case $T>T_c$, we here only show the result for the
former case. The solution for the scaling function takes the
form\cite{bariev}
\begin{mathletters}\label{bariev}
\begin{equation}\label{exact2}
{\cal F}(x,y)=\frac{y}{\sqrt{y^2+2}}\,\exp(-x/2)\,
\exp\left(\sum_{k=1}^{\infty}\frac{1}{k}\,f_k(x,y)\right)\>,
\end{equation}
with
\begin{eqnarray}\label{exact3}
\lefteqn{f_k(x,y)=-\frac{1}{\pi^k}\>\int_0^{\infty}dw_1\ldots
\int_0^{\infty}dw_k\>\>
\prod_{l=1}^{k} \,\exp\left[-x\,\mbox{ch}(w_l)\right]\> \times}\nonumber\\
& &\hspace*{4cm}
\frac{\mbox{ch}(w_l)-1}{\mbox{ch}(w_l)+\mbox{ch}(w_{l+1})}\,
\frac{\mbox{ch}(w_l)-1-y^2}{\mbox{ch}(w_l)+1+y^2}\>,\qquad w_{k+1}=w_1\>,
\end{eqnarray}
\end{mathletters}
where we abbreviated $\mbox{ch} \equiv \cosh$.

Several explicit asymptotic solutions
were given in Ref.\,\cite{bariev}. The case
$z<l_1<\xi$ was not considered, however. Also we ourselfs were not able
to obtain an explicit solution in the limit where $\xi$ goes to
infinity by $z/l_1$ remain at some finite value, the
case, where the anomalous short-distance behavior
discussed in Sec.\,\ref{twotwo} shows
up most clearly.
The so far only available explicit exact solution that is
consistent with our findings
is given by Eq.\,(28) of Ref.\,\cite{bariev}. It says that,
asymptotically for $x\ll 1$ and in the limit $y\to 0^+$, the
scaling function (\ref{exact2}) is given by
\begin{equation}\label{asympt}
{\cal F}(x,0^+)= E_3\,\left|\,\mbox{ln}\left(x/4\right)+1
\right|\,x^{3/8}
\left[1+\frac{3}{4}x+{\cal O}(x^2\,\mbox{ln}\,x)\right]
\end{equation}
with the constant $E_3=0.6987\ldots$

The asymptotic result (\ref{asympt})
is in agreement with our scaling analysis, which
itself provides a deeper understanding of the functional
form of (\ref{asympt}).
Firstly, the
exponent $3/8$ occurs and can be attributed to the
difference of scaling dimension as discussed in Sec.\,\ref{twotwo}.
Secondly, a logarithmic factor shows up, now
with an argument $\sim z/\xi$ instead of $\sim z/l_1$, however.
That effectively $\xi$ replaces $l_1$ for $l_1\to\infty$
can be understood on the basis
of our scaling analysis and the results reported in
Appendix\,\ref{appa}, since, as Eq.\,(\ref{m1tau}) tells us,
there is no logarithmic dependence of $m_1$ on $h_1$
for $h_1\to 0$ and finite $\xi$.

In order to obtain profiles for the whole
crossover region between the fixed points, one has to
compute the multiple integrals occuring in
(\ref{exact3}) numerically
to some finite order in the series.
The series converges rather quickly
such that in most cases the termination after the third term
is sufficient. Only very close to the surface
corrections due to higher orders become appreciable.

We evaluated (\ref{bariev}) numerically
for $K=0.999\,K_c$. The corresponding correlation
length is $\xi=567$ (lattice spacings). In Fig.\,9 the results for
the order-parameter profiles with the same values of $h_1$
as used in Figs.\,4 and 5 are displayed in
double-logarithmic representation ($z\equiv n-1$). Again the power
law $\sim z^{-1/8}$ is displayed for comparison. The resemblance
between the MC results shown in Fig.\,5 and the curves
of Fig.\,9 is
striking. One has to bear in mind, however, that
in the former case the crossover to the exponential
decay was caused by finite-size effects. In Fig.\,9, on
the other hand,
it is attributable to the finite correlation length.\\[2mm]

\def\epsfsize#1#2{0.6#1}
\hspace*{2.5cm}\epsfbox{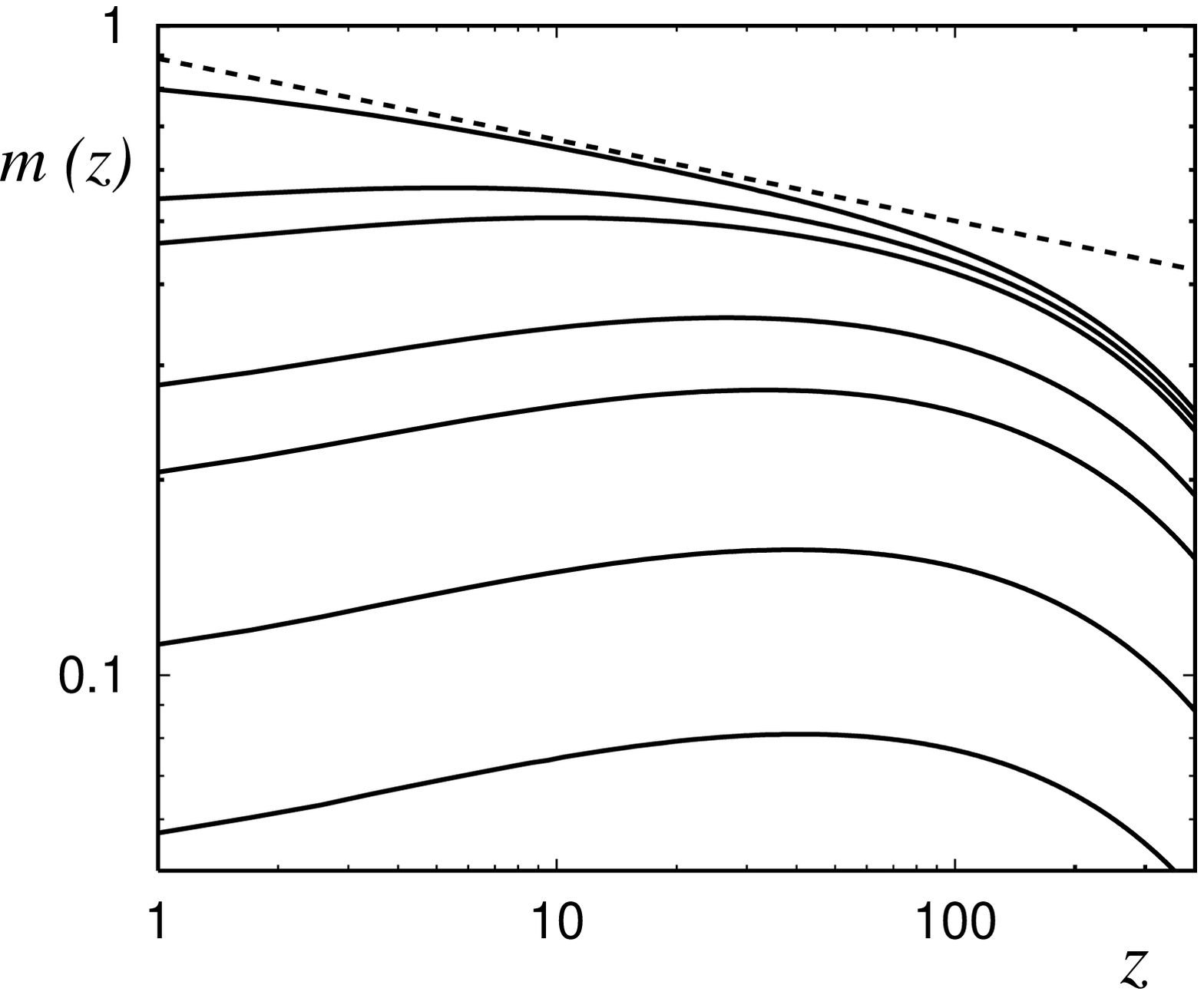}\\[0mm]
{\small {\bf Fig.\,9}: Order-parameter profiles
calculated from Bariev's solution
Eq.\,(\ref{bariev}) for the semi-infinite
2-$d$ Ising model, with
$K/K_c=0.999$, and for the same values of $h_1$ as
in Fig.\,4.
} \\[0.1cm]


\begin{thebibliography} {99}
\bibitem{foot} In the analysis in Ref.\,\cite{czeri} the surface exponents
were additionally labelled with the superscript ``ord" in order to
distinguish exponents of the various surface universality classes existing
in $d=3$. Since in the present work we focus on $d=2$, this distinction
is not necessary, the surface exponents appearing all belong to the
universality class of the ordinary transition.
\bibitem{foot5} Below the upper critical dimension $d^*$,
the scaling relation $y_1+x_1=d-1$ between the
operator dimensions holds. In the 2-$d$ Ising model $y_1=1/2$
such that $x_1$ takes the same value, $x_1=1-y_1$.
In general $x_1$ and $y_1$ have different values.
\bibitem{foot4} In Ref.\,\cite{bariev} a slightly more general system
with different coupling strenghts in directions parallel
and perpendicular to the surface was examined. We restrict our
presentation to the isotropic case.
\end{thebibliography}
\end{document}